\documentclass{article}
\usepackage{fortschritte}
\usepackage{amssymb,latexsym}
\begin {document}
\begin{flushright}
\hfill{HU-EP-03/71} \\
\hfill{hep-th/0310021}
\end{flushright}
                 
\def\titleline{
%
%
%
Heterotic String Theory on non--K\"ahler Manifolds  \\
\vspace{0.2cm}
with $H$--Flux 
and Gaugino Condensate
}
\def\email_speaker{
{\tt 
%
%
gcardoso,curio,dallagat,lust@physik.hu-berlin.de 
} \\ Proceedings of the 36th Symposium Ahrenshoop 
}
\def\authors{
%
%
%
%
%
G. L. Cardoso, G. Curio, 
G. Dall'Agata  and D. L\"ust\1ad\sp 
}
\def\addresses{
%
%
%
%
\1ad                                        %
Humboldt Universit\"at zu Berlin,
Institut f\"ur Physik,\\
Newtonstrasse 15, 12489 Berlin,
Germany %
}
\def\abstracttext{
%
%
We discuss compactifications of 
heterotic string 
theory to four dimensions in the presence of $H$--fluxes, which
deform the geometry of the internal manifold, and a gaugino condensate
which breaks supersymmetry.  
We focus on the compensation of the two effects in order to obtain
vacua with zero cosmological constant and we comment on the effective
superpotential describing these vacua. 

}
\large
\makefront
\section{Introduction}

String compactification in the presence of fluxes has been revived
recently as an appealing way to address the moduli problem.  
Turning on fluxes in the ten--dimensional string theories produces, at
the level of the effective four--dimensional action, a potential.  
Upon minimization of this potential one finds new vacua with
(generically) less moduli.  
Moreover, one expects to find Minkowski vacua only when the deformation
of the internal manifold balances the presence of the fluxes.

Supersymmetric compactifications of the heterotic string in the presence of
three--form fluxes need non--K\"ahler six--manifolds described in
\cite{Strominger:1986uh,Cardoso:2002hd,Becker:2003yv,Gauntlett:2003cy}.
More precisely one finds that a non--vanishing flux is associated to a
non--zero exterior derivative of the complex structure $J$ 
\cite{Gauntlett:2001ur}
\begin{equation}
\label{H-locking}
H = -\frac{1}{2} {\rm e}^{-8\phi} \, \star \,d({\rm e}^{8\phi}J).
\end{equation} 
This condition, which follows from the analysis of the
supersymmetry rules, can also be understood from the perspective of
rewriting the ten--dimensional effective
action of heterotic string theory
as a sum of squares \cite{CCDL2}.  The relevant term
for such a purpose is given by
\begin{equation}
\label{susy H square}
S=\frac{1}{2}\int {\rm e}^{8\phi}\,\left[H+\frac{1}{2} {\rm e}^{-8\phi} 
\, \star 
\,d({\rm e}^{8\phi}J)\right]^2\,.
\end{equation}
In this way it is clear that in
order to obtain a vanishing four--dimensional effective potential $V =
0$, the condition (\ref{H-locking}) has to be imposed.
Therefore, one obtains supersymmetric Minkowski vacua by fixing the
$(2,1)+(1,2)$ Hodge
components of the three--form flux, $H^{(2,1)}$ and 
$H^{(1,2)}$, in terms of the
internal geometry.  The other Hodge components, i.e. $H^{(3,0)}$ and
$H^{(0,3)}$, must vanish in order to have unbroken supersymmetry.

There is another but different physical effect in heterotic string theory 
which is naturally
connected with the appearance of an $H$--flux, namely gaugino condensation.
It is indeed known \cite{DRSW} that an interesting mechanism leading
to supersymmetry breaking while preserving a zero cosmological constant at the 
tree--level uses both a gaugino condensate and a non--vanishing flux.  
More precisely, one finds the following terms in the action,
\begin{equation}
\label{gluino H square}
S=\frac{1}{2}\int {\rm
e}^{8\phi}\, \left[H-\alpha^\prime 
(\bar{\chi}^A\Gamma_{(3)}\chi^A) \right]^2,
\end{equation}
(where $\Gamma_{(3)}= 1/3!\, e^{a} e^{b}e^{c}\, \Gamma_{abc}$) and assigning
an appropriate expectation value to ${\rm Tr} \, {\bar \chi} \Gamma_{mnp} \chi$
yields
\begin{equation}
\Sigma_{mnp}\equiv <{\rm Tr} \, {\bar \chi} \Gamma_{mnp} \chi > \rightarrow
\Lambda^3 \; \Omega_{mnp} +{\rm c.c.}
\;\;\;,\;\;\; \Lambda^3 = \frac14 < {\bar \lambda} (1 + \gamma_5)
\lambda > \;,
\label{gauc}
\end{equation}
where $\Omega$ is a  $(3,0)$--form on the internal manifold 
and $\Lambda^{3}$ denotes the expectation value of the gaugino
condensate in the four--dimensional space--time.  
Minkowski vacua follow then for Calabi--Yau compactifications if 
\begin{equation}
\label{H-gaug}
H = \alpha^\prime \left( \Lambda^3 \,\Omega + {\bar \Lambda}^3 \,
\bar \Omega\right).
\end{equation} 
Therefore, one obtains flat vacua breaking supersymmetry by fixing
the $(3,0)+(0,3)$ components of the three--form flux.

It is now natural to ask whether these two balancing mechanisms between
the flux and the geometry on one side and  the fermion
condensate on the other can be combined.
Since the two effects talk to different Hodge sectors of the $H$--flux,
it is natural to expect that they can be combined in the action in a
unique square,
\begin{equation}
\label{totsquare}
S = 
\frac12 \int  {\rm e}^{8\phi}\,
\left[H - \alpha^\prime \Sigma
+ \frac12 \star {\rm e}^{-8\phi}\, d({\rm e}^{8\phi}\, J)\right]^{2}\,.
\label{potgaug}
\end{equation}
In the following we will first discuss to what extent (\ref{totsquare})
is valid and then analyze what is the modification of the
superpotential needed in order to describe these vacua.

\section{Gaugino condensate, fluxes and torsion}

The bosonic part of the Lagrangean up to second order in
$\alpha^\prime$, including the gaugino condensate $\Sigma$,
is given by \cite{Bergshoeff:1989de}
\begin{eqnarray}
S &=& \int d^{10}x \, \sqrt{g}\, {\rm e}^{8 \phi}
\left[\frac14 \, R -\frac{1}{12}
\left(H_{MNP} - \alpha^\prime \,
\Sigma_{MNP} \right)^2 \,
+ 16 (\partial_M \phi)^2 \right. \nonumber\\
&&\qquad \qquad \qquad \;\;  \left. - \frac14  \alpha^\prime
\left(F_{MN}^{I} F^{I\,MN}-  R^+_{MNPQ} R^{+\,MNPQ}  \right)
\,\right]\,.
\label{action}
\end{eqnarray}
This action is written in the string frame and its fermionic
completion makes it supersymmetric using the three--form Bianchi
identity given by
\begin{equation}
dH = \alpha^\prime \left( \hbox{tr} \,  R^{+} \wedge
R^{+}  -\hbox{tr} F \wedge F\right)\,,
\label{HBI0}
\end{equation}
where the curvature $R^{+}$ is the generalized Riemann curvature built
from the generalized connection $\nabla^{+}$ (i.e. from $\omega^{\pm} = \omega
\mp H$).
Note that since we work at first order in $\alpha^\prime$, corrections to $\nabla^{+}$ or  $\nabla^{-}$ by the gaugino condensate 
$\Sigma$ can be neglected. 
Also note that it is the combination $H - \alpha^\prime \Sigma$ that enters in
the kinetic term for $H$, whereas it is only $H$ that enters in
the lhs of the Bianchi identity.  This asymmetry will result in the presence of
an additional term $d\Sigma \wedge ({\rm e}^{8 \phi} J)$
in the BPS rewriting of action (\ref{action}).

In the search for a BPS rewriting of (\ref{action}), and in the same spirit as
in \cite{CCDL2},
we will assume that the
space is given by the warped product of four--dimensional Minkowski spacetime  
with an internal space admitting an $SU(3)$ structure.  
In order to consistently obtain that setting to zero the BPS--like
squares implies a solution to the equations of motion, we also impose
that the only degrees of freedom for the various fields are given by
expectation values on the internal space and are functions only of the
internal coordinates.  

To simplify the discussion we limit ourselves to the case with
dilaton and  warp factor identified, i.e. $\phi = \Delta$, but the
generalization of the following results is straightforward.
After various manipulations, the action (\ref{action}) can be written as
\begin{eqnarray}
S  = && \int d^4 x \, \sqrt{g_{4}}\, \left\{- \frac{1}{2} \int_{{\cal
M}_{6}} \, {\rm e}^{8\phi}\left(8 d\phi + \theta \right)\wedge \star \left(8 d\phi + \theta \right)
+\frac18 \int_{{\cal M}_{6}} {\rm e}^{8\phi}\, J \wedge
J \wedge {\hat R }^{ab}J_{ab} \right.\nonumber \\
& -&\frac14 \int d^{6}y \; \sqrt{g_6} \,{\rm e}^{8\phi}\,
N_{mn}{}^p \,g^{mq}g^{nr}g_{ps}\,N_{qr}{}^s
-\frac{\alpha^\prime}{2}\, \int_{{\cal
M}_{6}} d  \Sigma \wedge ({\rm e}^{8 \phi} J)
\,
\nonumber \\
&+&  \frac12 \int_{{\cal M}_{6}} {\rm e}^{8\phi}\,
\left(H - \alpha^\prime \Sigma
+ \frac12 \star {\rm e}^{-8\phi}\, d({\rm e}^{8\phi}\, J)\right)
\wedge \star \left(H - \alpha^\prime \Sigma + \frac12 \star
{\rm e}^{-8\phi}\, d({\rm e}^{8\phi}\, J)\right) \,\nonumber\\
&-&  \frac{\alpha^\prime}{2}\int d^{6}y\; \sqrt{g_6}\,
 {\rm e}^{8\phi}\, \left[\hbox{tr} (F^{(2,0)})^{2} +
{\rm tr}(F^{(0,2)})^{2} + \frac14\,\hbox{tr} (J^{mn}F_{mn})^{2}\right] \nonumber \\
 &+& \left. \frac{\alpha^\prime}{2} \int d^{6}y\;\sqrt{g_6}
\, {\rm e}^{8\phi}\, \left[\hbox{tr} (R^{+\,(2,0)})^{2} +
\,\hbox{tr} (R^{+\,(0,2)})^{2} + \frac14 \,\hbox{tr} (J^{mn}
 R^{+}_{mn})^{2}\right]\right\}\,.
\label{finalaction}
\end{eqnarray}
In this expression the traces are taken with respect to the fiber
indices $a,b,\ldots$, whereas the Hodge type refers to the base
indices $m,n,\ldots$ of the curvatures.
The other geometrical objects appearing in the above
expression are the Lee--form
\begin{equation}
\theta \equiv J \lrcorner dJ= \frac{3}{2} J^{mn} \,
\partial_{[m}J_{np]} \, dx^p\,,
\label{eqLee}
\end{equation}
the Nijenhuis tensor
\begin{equation}
N_{mn}{}^p = {J_m}^q \partial_{[q}J_{n]}{}^p -  {J_n}^q
\partial_{[q}J_{m]}{}^p\,,
\label{eq:Nijenhuis}
\end{equation}
and the generalized curvature $\hat R$, which is constructed using the
Bismut connection built from the standard Levi--Civita connection and a
totally antisymmetric torsion $T^B$ proportional to the complex structure,
\begin{equation}
T^B_{mnp} = \frac32 \,{J_m}^{q} {J_n}^{r}
{J_p}^{s}\partial_{[q}J_{rs]} = -\frac32 J_{[m}{}^{q} \nabla_{|q|} J_{np]}\,.
\label{eq:tor}
\end{equation}
The action (\ref{finalaction}) will now be used to find the conditions
determining the background geometry and the form of the condensate $\Sigma$
by demanding the vanishing of (\ref{finalaction}).
Setting the squares to zero yields
\begin{itemize}

\item
the vanishing of the Nijenhuis tensor
$$N^{m}{}_{np} = 0\;,$$

\item the vanishing of
some components of the generalized Riemann
curvature constructed from the $\nabla^{+}$ connection,
$$R^{+\,(2,0)}=R^{+\,(0,2)}=J^{mn}R^{+}_{mn} = 0,$$

\item the vanishing of
\begin{equation}
d\phi +\frac18 \, \theta =0\, \,,
\label{dfth}
\end{equation}

\item
the vanishing of
\begin{equation}
H - \alpha^\prime \Sigma
+ \frac12 \star {\rm e}^{-8\phi}\, d({\rm e}^{8\phi}\, J) =0\;,
\label{condhs}
\end{equation}
\item
the vanishing of
$$F^{\,(2,0)}=F^{\,(0,2)}=J^{mn}F_{mn} = 0.$$

\end{itemize}
The vanishing of the Nijenhuis tensor states that the internal
manifold is complex.
The conditions on the $R^{+}$ curvature can be translated into the
 requirement of $SU(3)$ holonomy for the $\nabla^{-}$
connection.
The proof requires the identity 
\begin{equation}
R^{+}_{ab\,cd} = R^{-}_{cd\,ab} - (dH)_{abcd}\,,
\label{eq:rpm}
\end{equation}
which relates the $R^{+}$ and $R^{-}$ curvatures with the base and
fiber indices swapped (again, terms proportional to $\Sigma$ can be
neglected because they are of higher order in $\alpha^\prime$).
Using this identity and the fact that $dH$ gives higher order terms
in $\alpha^\prime$ the conditions on the base indices of $R^{+}$ become
conditions on the $R^{-}$ fiber indices, to lowest order in $\alpha^\prime$,
\begin{equation}
R^{-\,(2,0)} = R^{-\,(0,2)}=J^{ab}R^{-}_{ab} = 0\,.
\label{eq:rm}
\end{equation}
These conditions precisely state that the generalized curvature $R^-$
is in the adjoint representation of $SU(3) \subset SO(6)$ and
therefore its holonomy group is contained in $SU(3)$.

The conditions in the gauge sector are that the gauge
field strength is of type $(1,1)$ and $J$ traceless.  

On a complex manifold, the condition (\ref{condhs}) yields
\begin{eqnarray}
H^{(2,1) + (1,2)} &=& -  \frac12 \star {\rm e}^{-8\phi}\,d({\rm e}^{8\phi}\,
J) =
\frac12 i ( \partial- {\bar \partial }) J \;,\nonumber\\
H^{(3,0) + (0,3)} &=& \alpha^\prime \; \Sigma \;,
 \label{locking}
\end{eqnarray}
where we also used (\ref{dfth}).

On the solution, $R^+ \wedge R^+$ and $F \wedge F$ are of type $(2,2)$.
Therefore, the Bianchi identity (\ref{HBI0}) implies that $(dH)^{(3,1)
+ (1,3)} =0$, and hence
\begin{equation}
d\Sigma  = 0 \longrightarrow \Sigma = \Lambda^3 
\, \Omega + {\bar \Lambda}^3 \,\bar \Omega \;,
\label{ds}
\end{equation}
where $\Omega$ is now a holomorphic
(3,0)--form. 

Finally, let us discuss the 
term $d
\Sigma \wedge ({\rm e}^{8 \phi} J) $ in the action
(\ref{finalaction}).  This term vanishes on the solution, since $\Sigma$
is closed (\ref{ds}).  
Let us now consider its variation.  We obtain
\begin{equation}
\delta \int_{{\cal
M}_{6}} d  \Sigma \wedge ({\rm e}^{8 \phi} J) =
\int_{{\cal
M}_{6}} d \Sigma \wedge \delta ({\rm e}^{8 \phi} J)
+
\int_{{\cal
M}_{6}} d \, \delta \Sigma\, \wedge ({\rm e}^{8 \phi} J) \;,
\label{vartop}
\end{equation}
where we assumed that $\delta$ and $d$ commute.
The first term, when evaluated on the solution, vanishes due to (\ref{ds}).
The second term, however, is more problematic, because
$d \delta \Sigma$ may contain a piece of
$(2,2)$--type for non complex variations, unless one can show that it
vanishes when evaluated on the solution. 
Though this may seem reasonable if variations of a closed form result in
closed forms, it is not clear to us that (\ref{vartop})
vanishes on the solution.  
Notice that for the subclass of Calabi--Yau compactifications, the second term
doesn't contribute, since then $d J =0$.

We conclude our discussion with some remarks about the possible
superpotential describing such vacua in the effective theory.
It has been argued that a candidate superpotential describing the ${\cal N}
= 1$ vacua of the heterotic theory in the presence of fluxes is given
by \cite{Becker2,CCDL2}
\begin{equation}
    \label{flux supo}
W = \int_{{\cal M}_6}  {\cal H} \wedge \Omega = \int_{{\cal M}_6}
\left(H + i dJ\right)\wedge \Omega\,.
\end{equation}
In the presence of a gaugino condensate, 
the supergravity potential obtained from (\ref{finalaction}) 
contains a new contribution (\ref{potgaug})
proportional to the gaugino condensate $\Sigma$.  Therefore
it is natural
to expect that such a contribution will be captured by a shift in the
superpotential $W$ in the following way,
\begin{equation}
W \longrightarrow W + \int_{{\cal M}_6} \Sigma \wedge \Omega\,.
\label{supgau}
\end{equation}

On the other hand, it is known \cite{DRSW} that 
in the four-dimensional effective ${\cal N} = 1$
theory the gaugino condensate 
produces a 
non--perturbative contribution to the effective
superpotential 
which schematically is given by
\begin{eqnarray}
\label{Weff}
W_{eff}=c+{\rm e}^{-S}\,,
\end{eqnarray}
where $S$ denotes the four-dimensional dilaton--axion field.
On a Calabi--Yau threefold, $c$ is related to the three--form flux $H$,
whereas ${\rm e}^{-S}$ arises from 
$\int_{{\cal M}_6} \Sigma \wedge \Omega \sim {\bar \Lambda}^3 
\sim {\rm e}^{-S}$.

Inspection of (\ref{supgau}) now suggests that the 
microscopic description of $W_{eff}$ should
be given by
\begin{equation}
W = \int_{{\cal M}_6} \left(H + i dJ + \Sigma \right) \wedge \Omega\,.    
\label{supomic}
\end{equation}
A comparison of (\ref{supomic}) with (\ref{Weff}) then schematically yields
\begin{equation}
c \sim \int_{{\cal M}_6} (H + i d J) \wedge \Omega \;\;\;,\;\;\;
{\rm e}^{-S} \sim \int_{{\cal M}_6} \Sigma \wedge \Omega\,.
\end{equation}
The potential of the ${\cal N}=1$ effective theory is given by
(neglecting D-terms)
\begin{eqnarray}
V = {\rm e}^K \left( |DW|^2 -3|W|^2 \right) \;. 
\end{eqnarray}
Inspection of the potential (\ref{potgaug}) indicates that the resulting
model is of the no-scale type.
Using a tree-level K\"ahler potential
$K=-3\log (T+\bar T) - \log (S+\bar S)$ one then finds that
\begin{eqnarray}
V = 
\frac{1}{(S+\bar S)(T+\bar T)^3} | - W + (S+\bar S) W_S |^2\,,
\label{poteff}
\end{eqnarray}
where $W_S = \partial_S W$.
This does not fully coincide with (\ref{potgaug}) due to the presence
of $( S + {\bar S})$ in (\ref{poteff}).
This discrepancy was already noticed in the case of Calabi-Yau threefold
compactifications with gaugino condensates (see the second reference 
in \cite{DRSW}).


\end{document}